\documentclass[cameraready]{Interspeech}

\usepackage[nolist, nohyperlinks]{acronym}
\usepackage{booktabs}
\usepackage{tabularx}

\usepackage{multirow}
\usepackage{graphicx}
\usepackage{makecell} %

\usepackage{array} %

\usepackage{caption}
\usepackage{xcolor}
\usepackage{flushend}
\usepackage{hyperref}

\usepackage{amssymb}

\newcommand{\prevarrow}{%
  \tikz[baseline=-0.55ex, x=1ex, y=1ex, line cap=round, line join=round]{
    \draw[->, line width=0.35pt] (0,1.1) -- (0,0) -- (1.35,0);
  }%
}

\title{Repurposing a Speech Classifier for Guided Diffusion-Based Speech Generation}

\author[affiliation={1}, orcid=0009-0006-7319-6413]{Rostislav}{Makarov} %
\author[affiliation={1}, orcid=0000-0002-8678-4699]{Timo}{Gerkmann}

\address{
    $^1$ University of Hamburg, Germany
}

\email{firstname.lastname@uni-hamburg.de}

\keywords{diffusion, score matching, speech generation, classifier guidance}

\usepackage{comment}

\newlength{\eqtopspace}
\newlength{\eqbottomspace}

\newcommand{\eqpre}{\vspace{\eqtopspace}}
\newcommand{\eqpost}{\vspace{\eqbottomspace}}

\begin{acronym}
\acro{egc}[EGC]{Energy-based Classifier and Generator}
\acro{ebm}[EBM]{Energy-Based Model}
\acro{jem}[JEM]{Joint Energy-based Model}
\acro{pdf}[PDF]{Probability Density Function}

\acro{gan}[GAN]{Generative Adversarial Network}
\acro{sde}[SDE]{Stochastic Differential Equation}
\acro{ode}[ODE]{Ordinary Differential Equation}

\acro{dsm}[DSM]{Denoising Score Matching}
\acro{nn}[NN]{Neural Network}

\acro{cg}[CG]{Classifier Guidance}
\end{acronym}

\AtBeginDocument{%

  \setlength{\textfloatsep}{5pt plus 1pt minus 1pt}

  \setlength{\floatsep}{3pt plus 1pt minus 1pt}

  \setlength{\intextsep}{4pt plus 1pt minus 1pt}

  \setlength{\dbltextfloatsep}{5pt plus 1pt minus 1pt}

  \setlength{\dblfloatsep}{4pt plus 1pt minus 1pt}

\setlength{\abovedisplayskip}{2pt plus 1pt minus 1pt}
\setlength{\belowdisplayskip}{2pt plus 1pt minus 1pt}
\setlength{\abovedisplayshortskip}{0pt plus 1pt minus 0pt}
\setlength{\belowdisplayshortskip}{0pt plus 1pt minus 0pt}
  \captionsetup{skip=2pt}

}

\begin{document}

\maketitle

\begin{abstract}
Classifier guidance is a way to control diffusion generation by using a noise-conditioned classifier to steer the sampling process toward a target class. One drawback of classifier guidance is that it requires two separately trained models: a classifier and a diffusion model. We therefore study a more compact alternative in which a conventionally trained speech classifier is repurposed as the backbone for diffusion generation. Starting from a frozen noise-conditioned classifier in log-Mel space, we attach a lightweight subnetwork that reuses intermediate classifier representations and train only this subnetwork under a Denoising Score Matching objective. Our work shows that a pretrained classifier can be repurposed for conditional generation, 
providing an appealing bridge between discriminative modeling and conditional speech synthesis resulting in high speech quality within a single-backbone model, with reduced memory footprint and computational cost.
\footnote{The project page: audio samples and code: \url{https://sp-uhh.github.io/classifier-to-diffusion/}.}

\end{abstract}

\vspace{-0.1em}

\section{Introduction}
Score-based diffusion models have become a strong framework for generative modeling by learning the score of progressively perturbed data distributions and reversing a noise process at inference time \cite{songscoresde}. In recent work, this framework has also been successfully applied to speech generation \cite{diffwave, wavegrad, gradtts, edmsound}. A common approach to conditional generation is \ac{cg} \cite{classguidance, guided_tts}, where a noise-conditioned classifier steers the reverse dynamics toward a target label using input gradients. However, \ac{cg} is expensive: it requires a separately trained classifier and evaluation of both models at each reverse sampling step.

Instead of maintaining separate generative and discriminative components, the so-called \ac{jem} reveals a hidden generative model inside a classifier by interpreting its logits as an unnormalized joint model $p(X,y)$ over inputs $X$ and labels $y$ \cite{jem}. Under this view, the standard predictive distribution $p(y|X)$ remains unchanged, but the same logits also define an unnormalized marginal model $p(X)$ over inputs. Since diffusion sampling only requires a score $\nabla_{X} \log p(X)$, it can be obtained directly by differentiation through the classifier and has been used as an approximation for the true score \cite{egc}. 
However, treating the classifier itself as an energy-based model over $X,y$ requires dealing with an intractable normalizing constant and is known to cause training instabilities \cite{jem}.
We therefore use the \ac{jem} view only to extract score-relevant intermediate signals from a frozen classifier, while learning the final score with \ac{dsm}.

Motivated by this, we investigate whether a conventionally trained noise-conditioned classifier already provides sufficient information for score-based generation. 
We keep the classifier parameters fixed and train a lightweight generative adapter on top of its feature maps, following subnetwork and frozen-backbone adaptation strategies \cite{subnetwork, controlnet, mercea2024time}. 
This yields a parameter-efficient alternative to the standard two-model conditional generation pipeline.

Our main contributions are: (1) we show that a conventionally trained noise-conditioned speech classifier can be repurposed for diffusion generation; (2) we propose a parameter-efficient adaptation strategy that keeps the classifier frozen and trains only a lightweight generative subnetwork on its intermediate representations; 
and (3) we validate the method on SC09 benchmark \cite{warden2018speechcommands, diffwave}, comparing with a standard diffusion U-Net \cite{ddpm} and open-source SC09 models with fewer trainable parameters and lower compute, including low-data / zero-shot ablations against classifier-guided U-Net.

\begin{figure*}[t]
    \vspace{-10pt}
    \centering
    \includegraphics[width=0.9\textwidth]{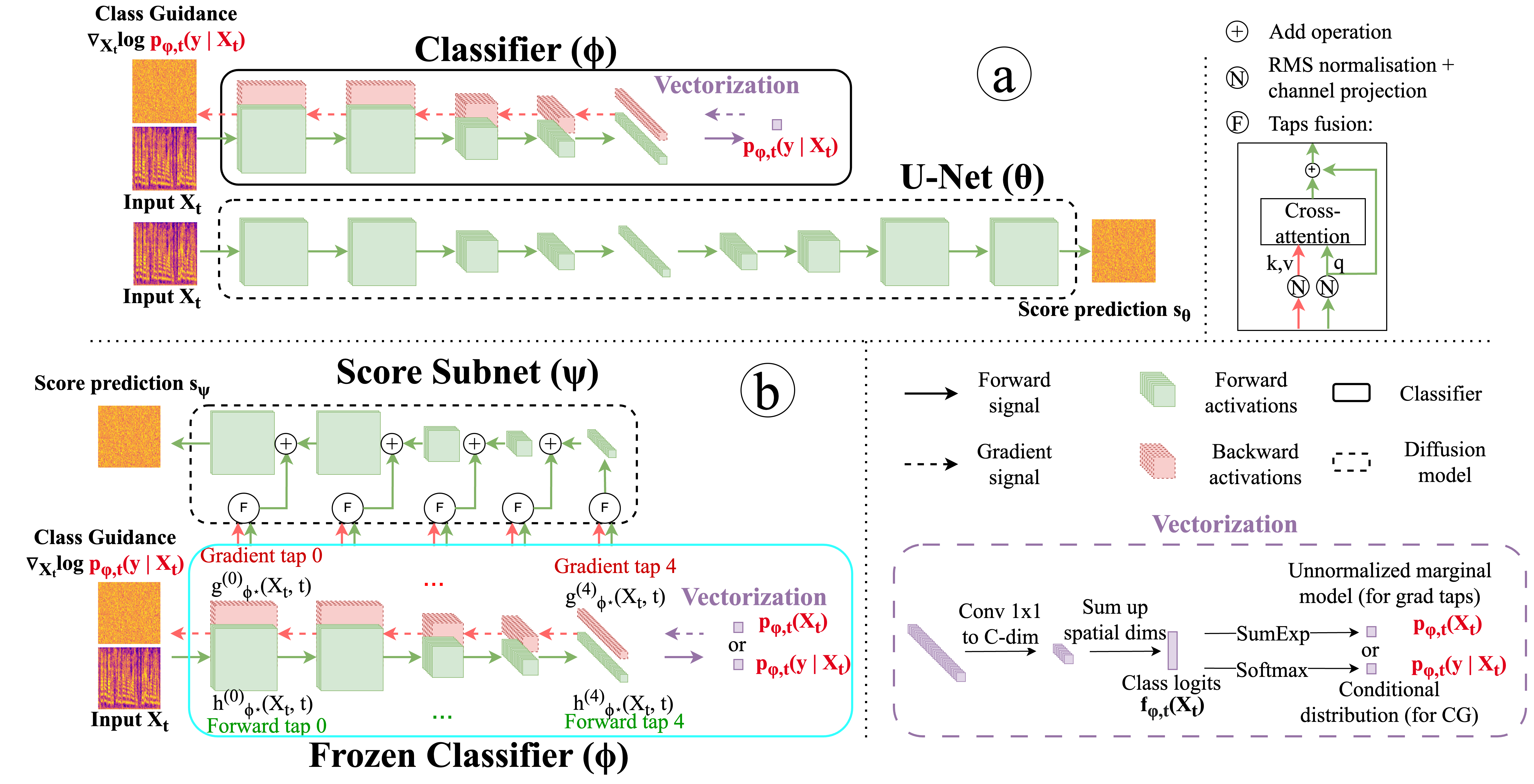}
    \caption{Conditional diffusion with classifier guidance. \textbf{(a)} Standard pipeline: a score model (U-Net) predicts the diffusion score, while a separate noise-conditioned classifier provides guidance by backpropagating the class probability $p_{\phi,t}(y|X_t)$ to the input.
     \textbf{(b)}~Score~Subnet: a frozen classifier backbone provides forward taps $h^{(k)}_{\phi^\star}$, while gradient taps $g^{(k)}_{\phi^\star}$ are obtained by backpropagating the JEM-style marginal probability $p_{\phi,t}(X_t)$ to the intermediate taps; the Subnet predicts the diffusion score from these signals.
    }
    
    \label{fig:subnet_scheme}
\end{figure*}

\section{Methodology}

\subsection{Score-Based Diffusion in Feature Space}
\label{subsec:diffusion}
\vspace{-0.35em}

We operate in a feature space of log-Mel filterbank features. Let $X_0 \in \mathbb{R}^{F \times N}$ denote a log-Mel spectrogram drawn from the data distribution $p_0$, where $F$ is the number of Mel frequency bins and $N$ is the number of time frames. Following \cite{songscoresde}, we define forward diffusion by means of the \ac{sde}

\eqpre \begin{equation}
    dX_t = f(X_t,t)\,dt + g(t)\,dW_t,\qquad t\in[0,1],
\end{equation} \eqpost

where $f(X_t,t)$ is the drift, $g(t)$ is the diffusion coefficient, $t$ is the continuous diffusion time, and $W_t$ is standard Brownian motion with the same dimensionality as $X_t$. Starting from $X_0 \sim p_0$, the forward \ac{sde} gradually adds noise and defines a sequence of time-indexed densities $\{p_t\}_{t\in[0,1]}$ for the random variable $X_t$, such that $p_1$ is close to a simple Gaussian distribution.
To generate samples, we draw $X_1 \sim p_1$ and then integrate the dynamics backward in time from $t=1$ to $t=0$. It has been shown that there exists a reverse \ac{sde} that reproduces the same distribution of $X_t$ for $t\in[0,1]$ as the forward process \cite{songscoresde}:

\eqpre \begin{equation}
    dX_t = \Big(f(X_t,t) - g(t)^2 \nabla_{X_t} \log p_t(X_t)\Big)\,dt + g(t)\,d\bar{W}_t.
    \label{eq:reverse_sde}
\end{equation} \eqpost

The term $\nabla_{X_t}\log p_t(X_t)$ is called the score function. We approximate it with a time-conditioned network $s_\theta$ such that:

\eqpre \begin{equation}
    s_\theta(X_t,t)\approx \nabla_{X_t} \log p_t(X_t)
    \label{eq:score_model}
\end{equation} \eqpost

We learn $s_\theta$ via \ac{dsm} \cite{songscoresde}, matching it to the perturbation score $\nabla_{X_t}\log q(X_t|X_0,t)$ available in closed form:

\eqpre \begin{equation}
    \begin{aligned}
    \mathcal{L}(\theta)
    &= \mathbb{E}_{t\sim  \mathcal{U}(0,1)}\,
    \mathbb{E}_{X_0\sim p_0}\,
    \mathbb{E}_{X_t\sim q(X_t|X_0,t)} \\
    &\quad \left[\left\Vert s_\theta(X_t,t)-\nabla_{X_t}\log q(X_t|X_0,t)\right\Vert_2^2\right].
    \end{aligned}
    \label{eq:dsm}
\end{equation}
In practice, we approximate statistical expectations $\mathbb{E}$ by means of empirical averaging. At inference time, we replace the unknown score in the reverse \ac{sde} with $s_\theta(X,t)$ and solve it numerically to generate samples.

\vspace{-0.35em}
\subsection{Classifier Guidance}
\label{subsec:cg}
\vspace{-0.35em}
For class-conditional generation we condition on $y$ (e.g., a class label or text prompt) and aim to sample $X_0\sim p(X|y)$. In score-based diffusion this requires the conditional score $\nabla_{X_t} \log p_t(X_t|y)$. By Bayes' rule,

\eqpre \begin{equation}
    \nabla_{X_t} \log p_t(X_t|y)=\nabla_{X_t} \log p_t(X_t)+\nabla_{X_t} \log p_t(y|X_t),
    \label{classifier_guidance}
\end{equation} \eqpost

since $\nabla_{X_t}\log p_t(y)=0$. As in \eqref{eq:score_model} we approximate the unconditional score $\nabla_{X_t}\log p_t(X_t)$ with a score model $s_\theta(X_t,t)$. The second term $\nabla_{X_t} \log p_t(y|X_t)$ is obtained by differentiating the log-probability of a noise-conditioned classifier $p_{\phi,t}(y|X_t)$ with respect to $X_t$, where the classifier is trained with cross-entropy on noisy inputs:

\eqpre \begin{equation}
    \mathcal{L}_{\mathrm{CE}}(\phi)=\mathbb{E}_{t,(X_0,y)}\,\mathbb{E}_{X_t\sim q(X_t|X_0,t)}
    \left[-\log p_{\phi,t}(y|X_t)\right].
    \label{eq:ce_noisy}
\end{equation} \eqpost

We can form a guided score

\eqpre \begin{equation}
    \tilde{s}(X_t,t,y) = s_\theta(X_t,t) + \gamma\,\nabla_{X_t} \log p_{\phi,t}(y|X_t),
\end{equation} \eqpost

where parameter $\gamma\ge 0$ controls the guidance strength. We then sample by plugging $\tilde{s}$ into the reverse \ac{sde} \eqref{eq:reverse_sde}.

\vspace{-0.35em}
\subsection{Joint Energy-based Model}
\label{subsec:jem_egc}
\vspace{-0.35em}
One prominent way to connect discriminative classifiers with generative models is to reinterpret classifier logits, $f_{\phi,t}(X_t)\in\mathbb{R}^C$ for $C$ classes, as unnormalized log-probability scores over classes. Following \ac{jem} \cite{jem} we can define these probabilities over input--label pairs $(X_t,y)$:

\eqpre \begin{equation}
    p_{\phi,t}(X_t,y)=\frac{\exp(f_{\phi,t}(X_t)[y])}{Z(\phi)} .
    \label{eq:jem_joint}
\end{equation} \eqpost

Here $f_{\phi,t}(X_t)[y]$ denotes the $y$-th component of the logit vector $f_{\phi,t}(X_t)$. The constant $Z(\phi)$ normalizes the distribution over $(X_t,y)$. Marginalizing out $y$ gives us a density model over inputs $X_t$:

\eqpre \begin{equation}
    p_{\phi,t}(X_t)=\sum_y p_{\phi,t}(X_t,y)=\frac{\sum_y \exp(f_{\phi,t}(X_t)[y])}{Z(\phi)} ,
    \label{eq:jem_marginal}
\end{equation} \eqpost

while dividing the joint distribution \eqref{eq:jem_joint} by the marginal \eqref{eq:jem_marginal} recovers the same classifier conditional distribution $p_{\phi,t}(y|X_t)$ used in Section~\ref{subsec:cg}.

In the score-based diffusion framework, sampling requires the true score $\nabla_{X_t}\log p_t(X_t)$. The \ac{jem} construction provides an analytic surrogate score by differentiating the classifier-based marginal \eqref{eq:jem_marginal}:

\eqpre \begin{equation}
    \begin{aligned}
    \nabla_{X_t}\log p_t(X_t)\ &\approx\ \nabla_{X_t}\log p_{\phi,t}(X_t) \\
    &=\ \nabla_{X_t}\log \sum_{y=1}^{C}\exp\left(f_{\phi,t}(X_t)[y]\right).
    \end{aligned}
    \label{eq:jem_score}
\end{equation} \eqpost

where the normalization constant cancels since $\nabla_{X_t}\log Z(\phi)=0$. 
Unlike \eqref{eq:score_model}, which learns the score with $s_\theta(X_t,t)$, the \ac{jem} formulation obtains a classifier-derived score by backpropagating the marginal log-density through the classifier. However, full JEM-style training involves an intractable normalization constant and has been reported to suffer from unstable optimization \cite{jem}. We therefore use this construction only at the representation level: forward activations and JEM-style activation gradients from a frozen classifier are used as inputs to a DSM-trained score subnet.

The \ac{egc} \cite{egc} applies this idea in diffusion by training a single time-conditioned classifier with both generative \eqref{eq:dsm} and discriminative \eqref{eq:ce_noisy} objectives.
However, \ac{egc} uses a time-conditioned U-Net classifier. The authors report markedly better generation quality for this architecture than for a standard feedforward ResNet, suggesting that strong performance in this framework depends on an encoder-decoder backbone \cite{egc}. This makes \ac{egc} approach less suitable for our goal of repurposing a classifier backbone, where we aim to retain a conventional classifier model.

\section{A Subnetwork for Score Prediction}
\label{subsec:subnetwork}

To repurpose a pretrained noise-conditioned classifier for diffusion generation without modifying its parameters, we attach a lightweight decoder-style adapter \emph{Score Subnet} to a frozen classifier backbone \cite{subnetwork} and train only the subnet with \ac{dsm} (Eq.~\eqref{eq:dsm}), reusing the classifier as a fixed multi-scale feature extractor instead of training a full encoder–decoder score model. Let $f_{\phi^\star,t}$ denote the frozen classifier. We extract intermediate feature maps \emph{taps} from stages $k=0,\ldots,K$,
$h^{(k)}_{\phi^\star}(X_t,t)$ (Fig.~\ref{fig:subnet_scheme}), taken from the same intermediate resolutions where a standard U-Net forms spatial skip connections. In addition, we compute \emph{gradient taps} by backpropagating the \ac{jem}-style marginal log-density from Eq.~\eqref{eq:jem_marginal} through the classifier:

\begin{equation}
    g^{(k)}_{\phi^\star}(X_t,t) =
    \frac{\partial \log p_{\phi^\star,t}(X_t)}{\partial h^{(k)}_{\phi^\star}(X_t,t)},
    \quad k=0,\ldots,K.
    \label{eq:bwd_taps}
\end{equation} \eqpost

This choice is motivated by the \ac{jem} score surrogate (Sec.~\ref{subsec:jem_egc}), which suggests that backpropagating classifier logits can reveal an unconditional, score-like signal, and that intermediate \emph{gradient taps} along this backpropagation path may provide useful information for the Score Subnet.
At each stage, we RMS-normalize the forward tap $h^{(k)}_{\phi^\star}$ and the gradient tap $g^{(k)}_{\phi^\star}$, project them to a common channel dimension, and fuse them via a cross-attention module.
After forming all fused taps, the decoder starts from the deepest one, applies several ResBlocks \cite{resnet}, upsamples to the next resolution, and then adds the following fused tap. This coarse-to-fine procedure is repeated until the original log-Mel resolution is reached.

We optimize only the subnet parameters $\psi$ under \ac{dsm}, keeping $\phi^\star$ fixed. At inference time, we use ${s}_\psi$ in the reverse \ac{sde} (Eq.~\eqref{eq:reverse_sde}) for unconditional sampling, and combine it with standard classifier guidance (Eq.~\eqref{classifier_guidance}) for conditional generation.

\begin{figure}[t]
    \centering
    \includegraphics[width=0.47\textwidth]{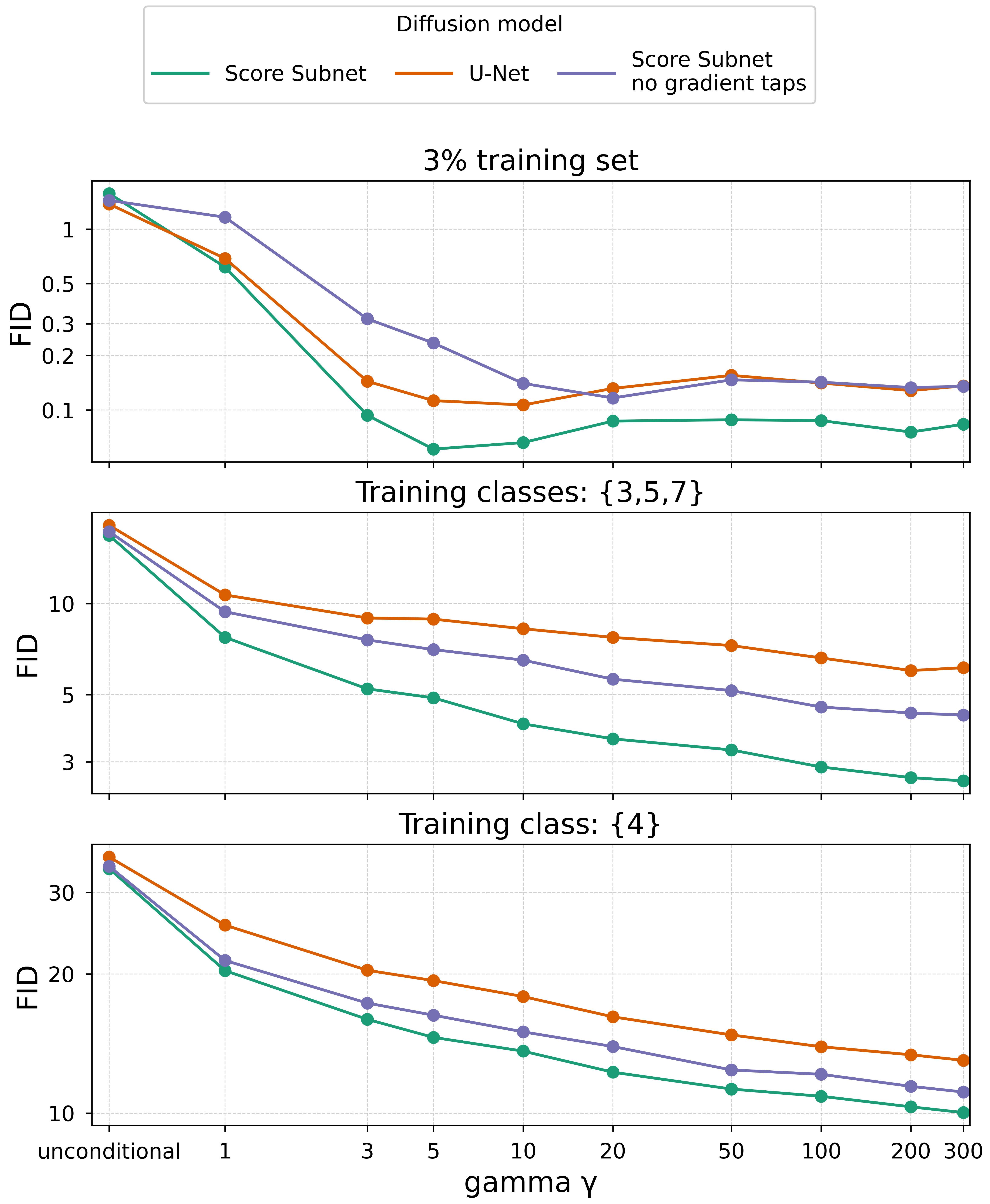}
    \caption{Effect of guidance strength $\gamma$ on SC09 conditional generation (FID$\downarrow$). We compare U-Net with Classifier Guidance sampling to our Score Subnet across three training regimes: 3\% training data, training label set $\{3,5,7\}$, and single-label training (label 4). The purple curve shows a Score Subnet variant without gradient taps. }

    \label{fig:gamma_plot}
\end{figure}

\section{Experimental Setup}
\label{sec:exp}
\vspace{-0.35em}
\subsection{Benchmark}
\vspace{-0.35em}
For training and evaluation, we use the class-labeled Speech Commands dataset \cite{warden2018speechcommands}, our primary benchmark is \textbf{SC09}, the spoken-digit subset (``zero''--``nine'') commonly used for speech generation \cite{diffwave, edmsound, sashimi}. We follow the official train/validation/test splits. 

\vspace{-0.35em}
\subsection{Speech features}
\label{subsec:speech_features}
\vspace{-0.35em}
We represent audio as log-Mel filterbanks using FFT size $1024$, hop length $256$, and $80$ Mel bins. We estimate global normalization statistics on the training split, subtract the training-set mean, and divide by the maximum absolute deviation to map features to $[-1,1]$. After the normalization, the feature standard deviation became below $0.5$, which stabilizes diffusion training. Since generation is performed in log-Mel filterbanks, we reconstruct waveforms using HiFi-GAN \cite{hifi} with a pretrained 16\,kHz vocoder\footnote{\href{https://huggingface.co/speechbrain/tts-hifigan-libritts-16kHz}{huggingface.co/speechbrain/tts-hifigan-libritts-16kHz}}.

\newcolumntype{L}[1]{>{\raggedright\arraybackslash}p{#1}}
\newcolumntype{C}[1]{>{\centering\arraybackslash}p{#1}}

\newcommand{\vgroup}[1]{%
  \rotatebox[origin=c]{90}{\footnotesize\bfseries\makecell[c]{#1}}%
}

\newcommand{\vsubgroup}[1]{%
  \rotatebox[origin=c]{90}{\scriptsize\makecell[c]{#1}}%
}

\begin{table*}[t]
\centering
\setlength{\tabcolsep}{3.2pt}      %
\renewcommand{\arraystretch}{0.93} %
\setlength{\extrarowheight}{0pt}   %

\caption{
Main results on SC09 for unconditional and classifier-guided conditional generation. We compare open-source baselines to a conventional U-Net score model and \emph{Score Subnet}, which predicts the diffusion score using a frozen classifier backbone. We report total and trainable parameters (\texttt{Total [Train]}) and sampling-time compute (GMACs per diffusion step). Conditional rows use classifier guidance with $\gamma=3.0$ and the same guidance classifier across methods, which Score Subnet reuses as part of its backbone. Metrics cover ScoreQ MOS, FAD, and ResNeXt-based fidelity/diversity. \textbf{Best} results within each section are highlighted.
}

\label{tab:main_results}
\small

\newcommand{\hdr}[1]{\textbf{#1}}
\newcommand{\hdrstack}[2]{\shortstack[c]{\hdr{#1}\\[-1.2pt]{\scriptsize #2}}}
\newcommand{\metr}[2]{\mbox{\hdr{#1}\,\ensuremath{#2}}} %
\newcommand{\hdrstackdown}[3][1.5ex]{%
  \raisebox{-#1}{\shortstack[c]{\hdr{#2}\\[-1.2pt]{\scriptsize #3}}}%
}

\begin{tabular*}{\textwidth}{@{\hspace{0.35cm}\extracolsep{\fill}}
L{3.55cm}  %
C{1.55cm}  %
C{1.05cm}  %
@{\hspace{5pt}\vrule width 0.5pt\hspace{5pt}} %
C{0.98cm}  %
C{0.98cm}  %
C{0.92cm}  %
C{0.92cm}  %
C{0.92cm}  %
C{0.92cm}  %
@{}}
\toprule
\hdr{Model} &
\hdrstackdown{Params (M)}{Total [Train]} &
\hdr{GMACs} &
\hdrstackdown{ScoreQ}{MOS $\uparrow$} &
\metr{FAD}{\downarrow} &
\multicolumn{4}{c}{\hdr{ResNeXt-based}} \\
\cmidrule(lr){6-9}
& & & & &
\metr{FID}{\downarrow} & \metr{IS}{\uparrow} & \metr{mIS}{\uparrow} & \metr{AM}{\downarrow} \\
\midrule

\multicolumn{9}{c}{\textbf{Unconditional generation}} \\
\addlinespace[1pt]

DiffWave~\cite{diffwave}\footnotemark      & 24.2          & -- & 2.85 & 2.59 & 1.92 & 5.26 &  51.21 & 0.68 \\
SaShiMi~\cite{sashimi}\footnotemark[\value{footnote}]       & 23.0          & -- & 2.79 & 1.84 & 1.42 & 5.94 &  69.17 & 0.59 \\
EDMSound~\cite{edmsound}     & 45.2          & -- & --   & --   & \textbf{0.14} & 7.17 & 160.2  & 0.33 \\

\cmidrule(lr){1-9}

U-Net & 16.6 [16.6] & 14.56 & {3.06} & \textbf{0.74} & {0.17} & {7.51} & {168.98} & {0.30} \\

Score Subnet (ours)  & {12.3 [4.4]} & {12.07} & \textbf{3.10} & {0.84} & {0.17} & \textbf{7.82} & \textbf{195.65} & \textbf{0.26} \\

\hspace{0.3em}\prevarrow\ no gradient taps & \textbf{11.9 [4.0]} & \textbf{7.71} & 3.04 & 0.90 & 0.28 & 7.02 & 126.95 & 0.37 \\

\midrule
\midrule

\multicolumn{8}{c}{\textbf{Conditional generation ($\gamma=3.0$)}} \\
\addlinespace[2pt]

U-Net + Classifier & 24.5 [16.6] & 22.74 & {3.25} & \textbf{0.82} & {0.03} & {8.36} & {260.50} & {0.19} \\

Score Subnet (ours) & {12.3 [4.4]} & {16.44} & \textbf{3.26} & 1.02 & {0.03} & \textbf{8.65} & \textbf{287.19} & \textbf{0.15} \\

\hspace{0.3em}\prevarrow\ no gradient taps & \textbf{11.9 [4.0]} & \textbf{11.50} & 3.23 & 0.94 & \textbf{0.02} & 8.05 & 223.98 & 0.22 \\

\bottomrule
\end{tabular*}

\end{table*}

\vspace{-0.35em}
\subsection{Architecture}
\label{subsec:arch}
\vspace{-0.35em}
\textbf{Backbone.} As a base architecture, we use a standard U-Net \cite{ddpm} with base width 64, three ResBlocks in the encoder, two ResBlocks in the bottleneck, and six ResBlocks in the decoder, with stride-2 convolutions for downsampling/upsampling.
Our \emph{classifier} is obtained from the U-Net by replacing its up path with a classifier head. The resulting model keeps the input projection, time embedding, down path, and bottleneck, and projects the flattened bottleneck representation to class logits.

\textbf{Score Subnet.} We attach the decoder-style Score Subnet described in Sec.~\ref{subsec:subnetwork} to the frozen classifier backbone. The subnet uses the same tap stages as the classifier encoder, starts from the deepest fused tap, and runs from coarse to fine resolution with upsampling--convolution modules and additive stage-wise merging. Each stage contains three ResBlocks with base width 32. A final GroupNorm--SiLU--Conv head maps the finest-resolution feature map to a one-channel score estimate ${s}_\psi(X_t,t)$ in log-Mel space.

\vspace{-0.35em}
\subsection{Training procedure}
\label{subsec:training_proc}
\vspace{-0.35em}
We train three components with all random seeds fixed:

\textbf{(i) U-Net diffusion baseline.}
We train a U-Net score model end-to-end with \ac{dsm} (Eq.~\eqref{eq:dsm}), directly predicting the score $s_\theta(X_t,t)$.

\textbf{(ii) Noise-conditioned classifier.}
We train the classifier encoder with cross-entropy on noisy inputs (Eq.~\eqref{eq:ce_noisy}) to obtain $p_{\phi,t}(y|X_t)$. This classifier is kept frozen during diffusion sampling via \ac{cg} and during subnet training.

\textbf{(iii) Score subnet on a frozen classifier.}
We freeze the trained classifier parameters $\phi^\star$ and train only subnet parameters $\psi$ with \ac{dsm} (Eq.~\eqref{eq:dsm}), using the intermediate representations \emph{taps} and \emph{gradient} taps as inputs (Sec.~\ref{subsec:subnetwork}). This yields a single-backbone generator that reuses the frozen classifier for intermediate representations.

\textbf{Sampling and guidance.}
All diffusion models use the variance-preserving (VP) \ac{sde} \cite{songscoresde} with 100 reverse-time steps and an Euler--Maruyama sampler. During sampling, we clip intermediate score predictions to $[-1,1]$ at each reverse step. For conditional generation, we apply classifier guidance (Eq.~\eqref{classifier_guidance}) using the frozen noise-conditioned classifier and sweep the guidance strength $\gamma$.

\footnotetext{Samples from \href{https://huggingface.co/krandiash/sashimi-release/}{{huggingface.co/krandiash/sashimi-release}}}

\vspace{-0.35em}
\subsection{Evaluation Metrics}
\vspace{-0.35em}
We use the standard SC09 benchmark metrics based on an auxiliary ResNeXt classifier, following prior work \cite{diffwave} and the protocol of the Unconditional Audio Generation Benchmark repository\footnote{\href{https://github.com/gzhu06/Unconditional-Audio-Generation-Benchmark}{{github.com/gzhu06/Unconditional-Audio-Generation-Benchmark}}}. These include FID (Fr\'echet distance in ResNeXt embeddings); Inception Score (IS), which rewards confident and diverse classifier predictions on generated samples; mIS, the same score computed in a class-balanced way; and Activation Maximization (AM), which reflects how well generated samples match the intended class under the auxiliary classifier.

Since these scores are tied to a particular pretrained ResNeXt model and use the training split as the reference, we additionally report two audio-specific metrics: Fr\'echet Audio Distance (FAD)\footnote{\href{https://github.com/gudgud96/frechet-audio-distance}{github.com/gudgud96/frechet-audio-distance}} computed from VGGish embeddings \cite{kilgour2019fad}, and SCOREQ, a non-intrusive speech-quality predictor that estimates MOS \cite{ragano2024scoreq}. We generate 2048 samples per model and evaluate FAD against the full SC09 test split (4107 utterances).

\textbf{Compute (GMACs).}
We report GMACs estimated with THOP\footnote{\href{https://github.com/Lyken17/pytorch-OpCounter}{github.com/Lyken17/pytorch-OpCounter}} for a one-second input and one reverse diffusion step. Forward pass MACs computed from \texttt{thop.profile}; backward MACs (for classifier) are approximated via autograd backward hooks on \texttt{Conv}/\texttt{Linear} layers, counting only input-gradient compute (\texttt{dL/dx}). We sum the relevant forward/backward terms for each inference path.

\section{Results}
In Tab.~\ref{tab:main_results} we present the main results on the SC09 benchmark, including parameter counts and GMACs.
For \textbf{unconditional} SC09 generation, we first report three open-source baselines (DiffWave, SaShiMi, EDMSound) together with parameter counts and three metric groups (speech quality, FAD, and ResNeXt-based). A U-Net trained end-to-end already reaches strong performance, outperforming the open-source baselines on perceptual quality and most ResNeXt-based metrics (e.g., {ScoreQ} $3.06$ and FAD $0.74$, with FID $0.17$), but uses a full $16.6$M-parameter generator and $14.56$ GMACs per step. Score Subnet matches the U-Net on FID $0.17$ and improves ScoreQ and the remaining ResNeXt-based metrics, while training only a $4.4$M subnet and reducing compute to $12.07$ GMACs. 
The \emph{no~gradient~taps} variant further reduces compute to about half of the U-Net baseline, but is worse than the full Score Subnet on most metrics, indicating that \emph{gradient taps} are beneficial.

For conditional generation ($\gamma{=}3.0$), Score Subnet performs on par with the standard classifier-guided U-Net pipeline, but is more compact and cheaper ($12.3$M total / $4.4$M trainable, $16.44$ GMACs vs.\ $24.5$M total, $22.74$ GMACs). Since the guidance classifier is part of the Score Subnet backbone, \ac{cg} reuses its forward pass. The \emph{no gradient taps} variant is cheaper still, reaching about half the U-Net+Classifier compute, but is slightly worse overall.

\textbf{Guidance study in low-data regimes.}
Fig.~\ref{fig:gamma_plot} reports FID on SC09 as a function of guidance strength $\gamma$ in low-data and zero-shot regimes. We train the guidance classifier on full SC09, then train the diffusion model or Score Subnet on restricted subsets (3\% data, labels $\{3,5,7\}$ only, or label 4 only), using the same frozen classifier for guided sampling. Across all settings, increasing $\gamma$ improves FID, and Score Subnet consistently outperforms the classifier-guided U-Net, with gains already visible at $\gamma \ge  3$. 
This suggests that the proposed method can reuse classifier representations for generation even when the score model sees only limited data or a subset of labels.

\section{Conclusions}
In this work, we show that a conventionally trained noise-conditioned speech classifier can be repurposed for diffusion-based speech generation in log-Mel space. By freezing the classifier and training only a lightweight Score Subnet on its intermediate representations, we obtain a compact generator that reduces both trainable parameters and compute compared to a standard U-Net and the standard two-model classifier-guidance. %
On the SC09 benchmark, the Score Subnet matches or slightly improves the strong U-Net and open-source baselines at a much lower cost. Furthermore, in low-data and zero-shot guidance regimes it consistently outperforms the classifier-guided U-Net.

\newpage
\section{Acknowledgements}
Funded by the Deutsche Forschungsgemeinschaft (DFG, German Research Foundation) -- 545210893, 498394658. The authors gratefully acknowledge the scientific support and HPC resources provided by the Erlangen National High Performance Computing Center (NHR@FAU) of the Friedrich-Alexander-Universität Erlangen-Nürnberg (FAU) under the NHR project f102ac. NHR funding is provided by federal and Bavarian state authorities. NHR@FAU hardware is partially funded by the German Research Foundation (DFG) – 440719683.

\section{Generative AI Use Disclosure}
Generative AI tools were used only for minor language editing (clarity, grammar, and polishing). All ideas, methods, experiments, results, and conclusions were produced and verified by the authors, who take full responsibility for the paper.

\bibliographystyle{IEEEtran}
\bibliography{mybib}

\end{document}